\documentclass[aps,prc,twocolumn,superscriptaddress]{revtex4-2}
\usepackage{amssymb}
\usepackage{lipsum} 
\usepackage{slashed}
\usepackage{amsmath}
\usepackage{amsfonts}
\usepackage{xcolor}
\usepackage{placeins}
\usepackage{graphicx}
\usepackage{soul}
\usepackage[colorlinks=true,
            linkcolor=blue,
            citecolor=blue,
            urlcolor=blue]{hyperref}

\newcommand{\red}[1]{}
\newcommand{\be}{\begin{equation}}
\newcommand{\ee}{\end{equation}}

\newcommand{\bra}{\langle}
\newcommand{\ket}{\rangle}

\newcommand{\bea}{\begin{eqnarray}}
\newcommand{\eea}{\end{eqnarray}}

\newcommand{\rvec}{\ensuremath{\boldsymbol{r}}}

\def\({\left(} 
\def\){\right)}
\def\[{\left[} 
\def\]{\right]}

\newcommand{\bs}[1]{\ensuremath{\boldsymbol{#1}}}

\graphicspath{{/}{./plots/}}
\begin{document}
\title{Scaling Laws for Three-Body Nuclear Contacts}
\author{Raz Yankovich}
\affiliation{The Racah Institute of Physics, The Hebrew University, 
Jerusalem 9190401, Israel}
\author{Ehoud Pazy}
\affiliation{Department of Physics,  NRCN, P.O.B. 9001, Beer-Sheva 84190, Israel}
\author{Nir Barnea}
\affiliation{The Racah Institute of Physics, The Hebrew University, 
Jerusalem 9190401, Israel}

\begin{abstract}
Three-nucleon short-range correlations (3N-SRCs) represent one of the least understood manifestations of short-range nuclear dynamics. We investigate these correlations within the generalized contact formalism and compute three-body nuclear contacts using a mean-field description of the long-range component of the nuclear wave function. 
These contacts quantify the probability of finding correlated nucleon triplets at short distances and provide a natural extension of the contact formalism beyond nucleon pairs. We find that the $^{3}$He and $^{3}$H contacts exhibit significant isospin-symmetry breaking, analogous to that observed previously for two-body contacts. Motivated by the semi-empirical mass formula, we derive a simple scaling relation for three-body contacts and show that it accurately reproduces the calculated values across medium-mass and heavy nuclei. Our results reveal a systematic dependence of 3N-SRCs on nuclear mass and composition, suggesting that three-body contacts obey universal scaling patterns closely analogous to those governing short-range-correlated nucleon pairs.
\end{abstract}
\maketitle
\section{Introduction}

Short-range correlations (SRCs) are a fundamental manifestation of the strong nuclear force at short distances. They arise when two or more nucleons approach one another within distances of approximately $1 \,\mathrm{fm}$, where the short-range repulsive core and tensor components of the nucleon--nucleon interaction become dominant~\cite{Hen:2016kwk,Ciofi15,Subedi08,HenSci14}. They generate the high-momentum components in the nuclear wave function, and play an important role in determining the structure and dynamics of nuclei~\cite{Carlson15,Hen:2016kwk}.

Over the past two decades, extensive experimental and theoretical efforts have established the existence of two-nucleon SRCs (2N-SRCs) as a universal feature of nuclear systems. Inclusive electron-scattering measurements at Bjorken $x_B>1$, exclusive nucleon-knockout experiments, and modern \emph{ab initio} calculations have demonstrated that high-momentum nucleons predominantly belong to strongly correlated nucleon pairs~\cite{Egiyan06,Subedi08,Fomin12,Hen2017}. A particularly striking observation is the universality of SRCs: after appropriate normalization, the high-momentum tails of nuclear momentum distributions exhibit nearly identical behavior across a broad range of nuclei. This universality suggests that short-distance nuclear dynamics are governed primarily by the underlying nucleon--nucleon interaction and are only weakly sensitive to the long-range structure of a specific nucleus.

The generalized contact formalism (GCF) provides a quantitative framework for describing this universality~\cite{Weiss15PRC,Weiss18}. Inspired by Tan's contact formalism for ultracold atomic systems~\cite{Tan08}, the GCF exploits the separation of scales between short-range and long-range nuclear dynamics. In this framework, short-distance observables are governed by nuclear contacts, which quantify the probability of finding correlated nucleon pairs with specific quantum numbers at short distances. The formalism successfully relates a wide range of observables, including momentum distributions, two-body densities, and electron-induced knockout reactions, and has been shown to reproduce both experimental measurements and \emph{ab initio} calculations~\cite{Weiss18,Weiss19,Torres21}.

While two-body SRCs are now relatively well understood, considerably less is known about three-nucleon short-range correlations (3N-SRCs). Three-body correlations probe regions of configuration space where three nucleons simultaneously approach one another and therefore provide a unique window into many-body aspects of the nuclear force. Understanding 3N-SRCs is particularly important because they are expected to be sensitive to three-nucleon interactions and may contribute to the short-distance structure of dense nuclear matter. Despite their importance, experimental evidence for 3N-SRCs remains limited and significantly less conclusive than for the two-body case. Inclusive scattering measurements have identified kinematic regions in which 3N-SRC contributions are expected to dominate, but a clear experimental isolation of a universal 3N-SRC regime has yet to be achieved~\cite{Fomin23}.

Recent developments have extended the contact formalism to three-body correlations, providing a natural theoretical framework for the study of 3N-SRCs~\cite{Weiss3body}. These advances raise a number of fundamental questions. How do three-body contacts depend on nuclear mass and composition? Do they exhibit simple scaling patterns analogous to those observed for two-body contacts? To what extent can their behavior be understood using the same scale-separation arguments that underlie the success of the GCF for nucleon pairs?
\par

In this work, we address some of these questions by calculating three-body nuclear contacts within the generalized contact formalism. We combine the universal short-range structure of the three-body wave function with a mean-field description of the long-range nuclear dynamics and evaluate contacts across a broad range of nuclei. We find that the resulting contacts exhibit systematic trends throughout the nuclear chart, including significant isospin-symmetry breaking between the $^{3}\mathrm{He}$ and $^{3}\mathrm{H}$ channels. Motivated by an analogy with the semi-empirical mass formula, we derive a simple scaling relation for three-body contacts and demonstrate that it accurately reproduces the calculated results across medium-mass and heavy nuclei. Our findings indicate that three-body short-range correlations obey remarkably simple and universal patterns, extending to three-body systems the scaling behavior previously observed for short-range correlated nucleon pairs.


\section{Methods} 
The GCF is based on the principle of scale separation, which enables, to a good approximation, the factorization of short-range observables from long-range nuclear dynamics. By separating low and high-momentum scales, this factorization provides a powerful framework for identifying relations between seemingly unrelated observables. It also allows many-body methods optimized for describing long-range nuclear structure to be employed in the calculation of short-range observables. The method has been verified for pairwise SRCs  \cite{YPB25} by comparing results with charge density measurements \cite{Weiss19}, and inclusive electron scattering cross-section ratios \cite{Fomin12,Fomin17}. 
\par
At its core, the GCF relies on a factorization ansatz for the nuclear many-body wave function. When examined at short interparticle distances, the wave function can be decomposed into a universal few-body component describing a cluster of close (correlated) nucleons, multiplied by a residual function that depends on the specific nucleus and quantum state. The latter encodes the motion of the remaining nucleons and can be described within a mean-field–like framework.
Thus, in the  extension of the GCF to the 3N-SRCs case, we consider the limit where three nucleons, say $ijk$  
are in close proximity, i.e. $r_{ij}\rightarrow 0$, $r_{ijk} \rightarrow 0$,
where $\rvec_{ij}=\rvec_j-\rvec_i$, and $\rvec_{ijk}=\rvec_k-(\rvec_i+\rvec_j)/2$
are the three-body Jacobi vectors.
In this limit we expect the many-body nuclear wave-function to factorize
in the following way \cite{Weiss3body}:
\be\label{eq:three_body}
  \Psi\longrightarrow
  \sum_\alpha\varphi_{abc}^\alpha\left(\rvec_{ij},\rvec_{ijk})
  B_{abc}^\alpha(\bs{R}_{ijk},\{\rvec_l\}_{l\not=i,j,k}\right),
\ee
where $\rvec_k$ are the single particle coordinates, and $\bs{R}_{ijk}=(\rvec_i+\rvec_j+\rvec_k)/3$ is the three-body center of mass vector.
$\varphi_{abc}^\alpha$ are the three-body universal functions, given by the 
zero-energy solutions of the three-body Schrödinger equation \cite{Saar23},
and $B_{abc}^\alpha$ are the "spectator" part of $\Psi$.
The index $\alpha$ defines the quantum numbers for the three-body states,
and $abc$ specifies the triplet type ($nnn$, $nnp$, $npp$ or $ppp$). 
Three-body SRCs are expected to be dominated by 
tritium ($^{3}\text{H}$) and helium-3 ($^{3}\text{He}$) like configurations,
whereas isospin $t=3/2$ configurations, such as $ppp$, and $nnn$ are expected to be suppressed due to the Pauli exclusion principle.
Hence, in the following we will use the short hand notations $C_{nnp}$, and $C_{npp}$ to denote the 
$^{3}\text{H}$, and $^{3}\text{He}$ contacts respectively.

Following the two-body example \cite{Weiss15PRC}, the three-body contact matrix is defined by 
\be\label{eq:contact3B}
   C_{abc}^{\alpha \beta} \equiv N_{abc} \bra B_{abc}^\alpha | B_{abc}^\beta \ket~
\ee 
 {where} $N_{abc}$ is the number of $abc$ triplets in the nucleus.
 { Notice that~\eqref{eq:contact3B} implies that the contact is a propery independent of the specific form of the short range universal wavefunction}. 
The three-body contact of nucleus $A$, $C_{abc}^{\alpha\beta}(A)$
can be determined, up to a constant normalization factor, 
by evaluating the expectation value of a short-range operator 
such as 
\be\label{eq:SRC_op}
\hat{O}_{abc}^{\alpha\beta} = \sum_{i<j<k}
   {\hat{\delta}(\rvec_{ij})\hat{\delta}(\rvec_{ijk})\hat{P}_{abc}^{\alpha\beta}}.
\ee
Here, the triplet type as well as the contact quantum numbers $\alpha\beta$ 
are determined by the projection operator $\hat{P}_{abc}^{\alpha\beta}$.
Utilizing the asymptotic form of the many-body wave function \eqref{eq:three_body},
the expectation value of $\hat{O}_{abc}^{\alpha\beta}$ given
a nuclear wave-function $|\Psi_A\ket$, reads
\be \label{eq:A_body_c}
  \bra \Psi_A | \hat{O}_{abc}^{\alpha\beta}| \Psi_A \ket 
  \cong   
  C_{abc}^{\alpha\beta}(A)
  \bra{\varphi_{abc}^{\alpha}}|
  \hat{\delta}(\rvec_{ij})\hat{\delta}(\rvec_{ijk})\hat{P}_{abc}^{\alpha\beta}
  |\varphi_{abc}^{\beta}\ket~.
\ee
Considering now the ratio of two such expectation values, associated
with two different nuclei, say $A$ and $B$, the universal
part is canceled out and one is left with
the contact ratio, 
\be\label{c_ratios}
   \frac{\bra \Psi_A|\hat{O}_{abc}^{\alpha\beta}|\Psi_A\ket}
        {\bra\Psi_{B}|\hat{O}_{abc}^{\alpha\beta}|\Psi_{B}\ket}
   =
   \frac{C_{abc}^{\alpha\beta}(A)}{C_{abc}^{\alpha\beta}(B)}.
\ee
Eq.~\eqref{c_ratios} implies that the ratio of expectation values of short range operator is independent of the specific form of the universal wave function describing the correlated triplet. It is equal to the contact ratio, which relies 
solely on the triplet type and its quantum numbers.
\par
Computing the contact demands a good approximation for the spectator function
$B_{abc}^\alpha$. As $B^{\alpha}_{abc}$ is a long wavelength wavefunction, an approximation may be achived by using mean field theory, as was done for pairs in \cite{YPB25}.

Specifically, as the contact ratio is independent of the short wavelength part of the wave-function, 
Eq. ~\eqref{c_ratios}, we assume that
\be \label{eq:shell}
   \frac{C_{abc}^{\alpha\beta}(A)}{C_{abc}^{\alpha\beta}(B)}
   =
   \frac{\bra \Psi_A|\hat{O}_{abc}^{\alpha\beta}|\Psi_A\ket}
        {\bra\Psi_{B}|\hat{O}_{abc}^{\alpha\beta}|\Psi_{B}\ket}
   \cong
   \frac{\bra \Phi_A|\hat{O}_{abc}^{\alpha\beta}|\Phi_A\ket}
        {\bra\Phi_{B}|\hat{O}_{abc}^{\alpha\beta}|\Phi_{B}\ket}~,
\ee
 {where $|\Phi_{A,B}\ket{}$ are the mean field approximations for the nuclear wavefunctions. The simplest form of $|\Phi_A \ket$ is a single Slater determinant of single particle states.} 
Various mean-field approximations of the many-body wave function can be systematically explored with increasing sophistication, such as
the Fermi gas model, the harmonic oscillator (HO) model, or the Woods-Saxon (WS) model.

Fermi gas calculations can be performed analytically. The resulting probablity of observing a 
2N-SRCs in a nuclear gas of density $\rho$ is 
$C^{\alpha\beta}_{NN}/A\propto \rho$, and for
3N-SRCs it is $C^{\alpha\beta}_{NNN}/A\propto \rho^2$. As the density of the atomic nucleus is
roughly a constant, these results suggest that both the two-body and the three-body contacts should 
grow linearly with the number of nucleons.

To gain further insight into the weight of three-body SRCs in the nuclear wave function,
one can introduce a geometric interpratation of SRCs, where the 
corrleated particles are
confined into a small volume $v_\text{src}$. Utilizing this interpratation and the Fermi gas model
the probablity of finding correlated two nucleon pairs is given by $C_{NN}/A = (3/8) \rho v_\text{src}$,
and the probablity of finding SRCd triplets is $C_{NNN}/A = (1/16) (\rho v_\text{src})^2$. 
As the weight of SRCs is about 20\% in the nuclear wave-function, we may conclude that
the weight of 3N-SRCs should be about an order of magnitude smaller.

Going beyond the Fermi gas, using more sophisticated models requires numerical treatment, see  \cite{Yankovich26}. 
In the following we present results obtained for the 3N-SRCs contacts obtained representing $|\Psi_A\ket$  as a single 
Slater determinant $|\Phi_A\ket$, with single-particle orbitals of  phenomenological WS potential.
We examine the model's sensitivity to the WS potential parameters by comparing two parameterizations: 
the so called universal parameterization (UR) \cite{Dudek82}, and the Schewierz, Wiedenhöver, and Volya (SWV) parameterization \cite{Schewierz13}. 

\section{Results} 

\begin{figure}
\centering
\includegraphics[scale=0.55]{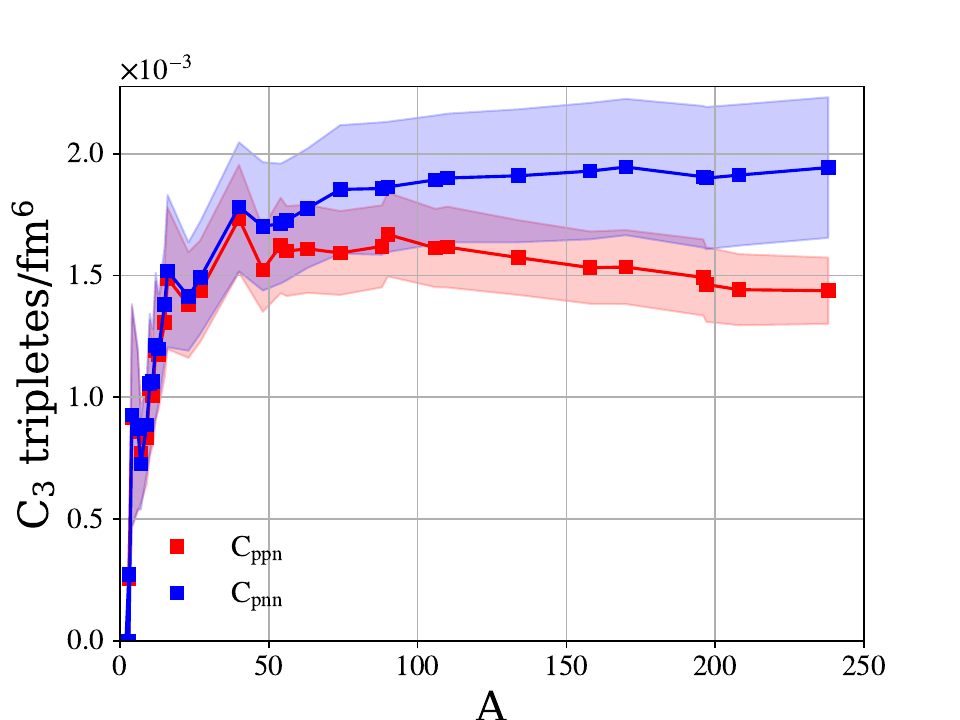}
\caption{ The two 3N-SRC contacts, corresponding to $nnp$ and $ppn$ configurations as a function of the $A$. The central results and error bands represent the average and difference of
the two WS paremetrizations - SWV and UR.}
\label{fig:3b_MF_res}
\end{figure}

Fig.~\ref{fig:3b_MF_res} presents the calculated three-body contacts for the $npp$ and $nnp$ channels obtained using 
Eq.~\eqref{eq:shell} and the appropriate SRC operators of Eq.~\eqref{eq:SRC_op}. As expected, both contacts increase 
approximately linearly with the nuclear mass number for $A\gtrsim 50$, indicating that the probability of finding short-range correlated triplets scales with the size of the nucleus. While the overall behavior is similar for the two channels, small but systematic differences are observed. In particular, the $npp$ contact exhibits a slightly sublinear dependence on $A$, whereas the $nnp$ contact displays a weak superlinear trend. A qualitatively similar behavior was previously reported for spin-zero two-body contacts~\cite{YPB25}, suggesting that analogous mechanisms govern the scaling of two- and three-body short-range correlations.

\begin{figure}
\centering
\includegraphics[scale=0.55]{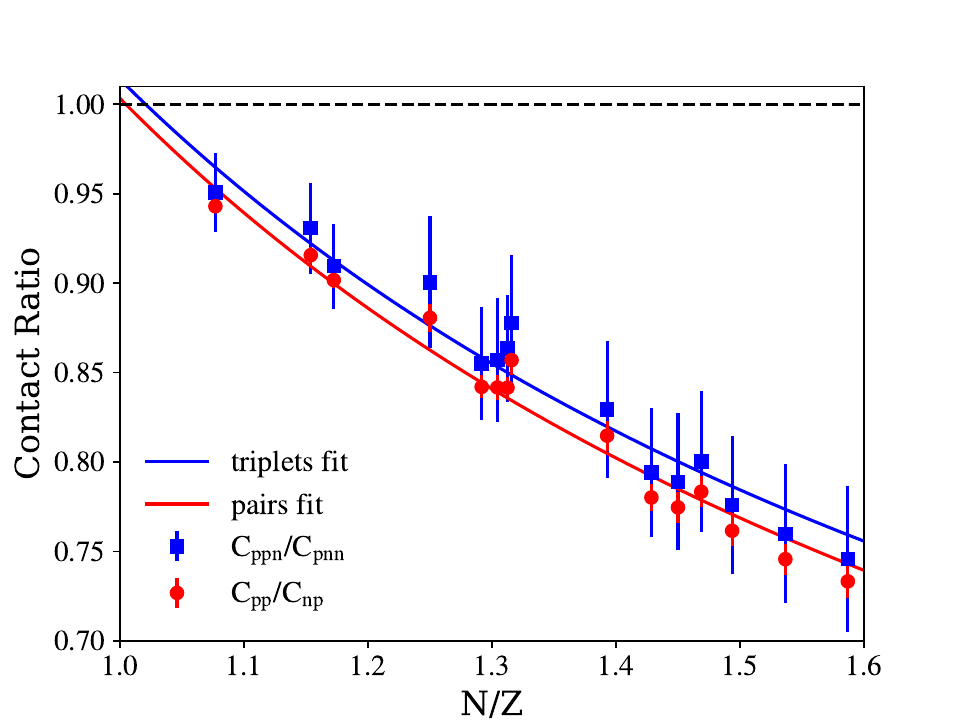}
\caption{ {The ratios ${C_{npp}/C_{nnp}}$ and {$C^{s=0}_{pp}/C^{S=0}_{np}$} as a function of neutron excess ${N/Z}$}}
\label{fig:NbvZ_ratio}
\end{figure}

To investigate this similarity further, we examine the ratio of the two three-body contacts,
${C_{npp}}/{C_{nnp}}$,
and compare it with the corresponding two-body ratio $C^{s=0}_{pp}/C^{s=0}_{np}$ reported in Ref.~\cite{YPB25}. 
The results are shown in Fig.~\ref{fig:NbvZ_ratio} as a function of the neutron excess $N/Z$. 
Following Ref.~\cite{YPB25}, the calculated ratios were fitted using the phenomenological expression
\begin{equation}\label{isospin_ratio}
  \frac{C_{npp}}{C_{nnp}} = 
  \alpha+\beta\left(\frac{Z}{N}-1\right).
\end{equation}
The fitted parameters are listed in Table~\ref{tab:NbvZ}. Remarkably, the extracted values of $\alpha$ and $\beta$ are statistically consistent with those obtained for the two-body contact ratio. This agreement suggests that the isospin dependence of three-body contacts is governed by the same underlying geometric and density effects that control the behavior of short-range correlated pairs.

The close correspondence between the scaling behavior of two- and three-body contacts motivates the search for a simple phenomenological description of the three-body contacts. In the following, we derive scaling relations for three-nucleon SRCs by extending the geometric arguments previously developed for two-body contacts~\cite{YPB25,Liang24}. The resulting expressions provide an intuitive interpretation of the mass and isospin dependence of the calculated contacts and allow a direct comparison with the mean-field results.

\begin{table}[h]
\centering
\caption{
Parameters of the fit
$ C_{npp}/C_{nnp} = \alpha+\beta(Z/N-1) $, Eq. \eqref{isospin_ratio},
and the analogous fit for the spin-zero two-body contact ratio
($C_{pp}^{s=0}/C_{np}^{s=0}$).
The close agreement between the fitted coefficients suggests that the
isospin dependence of two- and three-body short-range correlations is
governed by a common underlying mechanism.
}
\label{tab:NbvZ}
\begin{ruledtabular}
\begin{tabular}{lcc}
 & $C_{npp}/C_{nnp}$ & $C^{s=0}_{pp}/C^{s=0}_{np}$ \\
\hline
$\alpha$ & $1.01 \pm 0.02$ & $1.003 \pm 4\times10^{-4}$ \\
$\beta$ & $\ 0.69 \pm 0.09$ & $0.70 \pm 0.02$ \\
$\text{cov}(\alpha,\beta)$  & $2\times 10^{-3}$ & $7\times 10^{-5}$
\end{tabular}
\end{ruledtabular}
\end{table}

\subsection{Scaling laws}

Following the approach developed for two-body contacts \cite{YPB25}, we derive 
a scaling relation for three-body contacts based on simple geometric 
considerations. We begin by estimating the probability of finding three 
nucleons within a short-range correlation volume $v_{abc}$.

Considering a nucleon of type $a$ located in a small volume element $\Delta V$, and assuming locally uniform nuclear density,
the probability of finding two additional nucleons of types $b$ and $c$ within 
the correlation volume $v_{abc}$ surrounding nucleon $a$ is
\be
  P(bc|a) =
  N_b\frac{v_{abc}}{\Delta V}
  N_c\frac{v_{abc}}{\Delta V}
  \propto
  \rho_b \rho_c v_{abc}^{\,2},
\ee
where $N_b$ ($N_c$) and $\rho_b$ ($\rho_c$) denote the number and density of nucleons of type $b$ ($c$), respectively. 
If any of the nucleons types are identical, the corresponding combinatorial factors are modified accordingly, e.g. $N_b=N_a-1$ when $a=b$.

The density of correlated triplets is therefore
\be\label{eq:GenerlaizedL}
  \rho_{abc}(A) = \rho_a\rho_b\rho_c\,v_{abc}^{\,2}
  \propto
  \frac{N_a}{R_a^3}
  \frac{N_b}{R_b^3}
  \frac{N_c}{R_c^3}
  v_{abc}^{\,2},
\ee
where $R_a$ denotes the radius associated with nucleons of type $a$, and same for $b,c$.

Defining
\be
  L_V^{abc}\equiv v_{abc}^{\,2},
\ee
the volume contribution to the three-body contact can be written as
\be\label{eq:C3NV}
  C_{abc}^{V} \approx 
  L_V^{abc}\int \rho_a\rho_b\rho_c\, d\Omega_{abc}
  \propto
  L_V^{abc}
  \frac{N_a}{R_a^3}
  \frac{N_b}{R_b^3}
  \frac{N_c}{R_c^3}
  \Omega_{abc},
\ee
where $\Omega_{abc}$ denotes the volume available for forming a correlated triplet of type $abc$. For triplets composed of a single nucleon species, $\Omega_{abc}\propto R_a^3$, whereas for mixed triplets such as $nnp$ and $npp$, $\Omega_{abc}\propto R^3$, where $R$ is the nuclear matter radius. 

The volume term alone neglects finite-size effects associated with nucleons near the nuclear surface. In analogy with the surface contribution of the semi-empirical mass formula, we introduce a correction proportional to the nuclear surface area. The basic density estimate of Eq.~(\ref{eq:GenerlaizedL}) remains unchanged, but the integration is restricted to a surface layer of thickness $\Delta R$,
\be\label{eq:C3NS}
  C_{abc}^{S} \approx
  \int dS\,\Delta R\,\rho_a\rho_b\rho_c\,v_{abc}^{\,2}
  \propto
  \frac{N_a}{R_a^3}
  \frac{N_b}{R_b^3}
  \frac{N_c}{R_c^3}
  L_S^{abc}S_{abc},
\ee
where
\be
   L_S^{abc}\equiv \Delta R\,v_{abc}^{\,2},
\ee
and $S_{abc}\propto R_{abc}^2$ is the relevant surface area.

Combining the volume and surface contributions yields the three-body scaling relation
\be\label{eq:C3N_scaling}
  C_{abc} \approx
  \frac{N_a}{R_a^3} \frac{N_b}{R_b^3} \frac{N_c}{R_c^3}
  \left( L_V^{abc}R_{abc}^{3} + L_S^{abc}R_{abc}^{2} \right).
\ee

For comparison, and following Ref.~\cite{YPB25}, we introduce the analogous expression for the two-body contacts,
\be\label{eq:C2N_scaling}
  C_{ab} \approx
  \frac{N_a}{R_a^3} \frac{N_b}{R_b^3} \left(L_V^{ab}R_{ab}^{3} + L_S^{ab}R_{ab}^{2} \right),
\ee
where $L_V^{ab}=v_{ab}$, and $L_S^{ab} = \Delta R v_{ab}$, with $v_{ab}$ being the two-body correlation volume. 

\subsection{Validation and isospin symmetry}

The scaling relations given by Eqs.~(\ref{eq:C3N_scaling}) and
(\ref{eq:C2N_scaling}) were fitted to the contacts obtained from the
Woods--Saxon mean-field calculations. Separate fits were performed for
the UR and SWV parameterizations. In each case we considered both a
two-parameter fit, with $L_V$ and $L_S$ treated as free parameters, and
a constrained fit with $L_S=0$.

\begin{figure}
\centering
\includegraphics[scale=0.55]{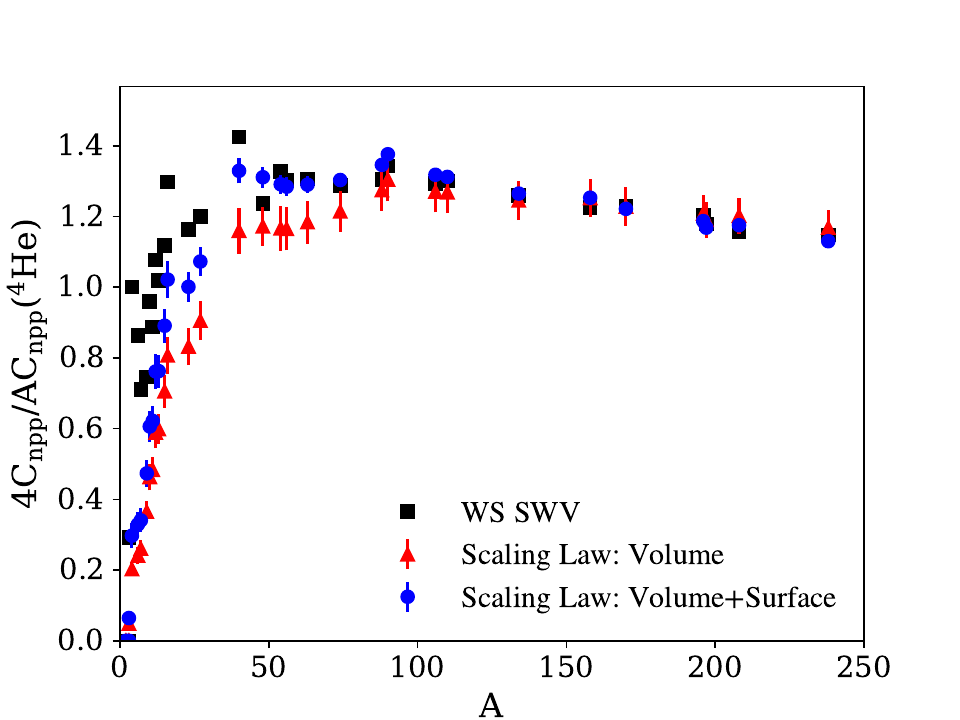}
\caption{
Mass-number dependence of the normalized $npp$ three-body contact, 
$4 C_{npp}(A)/A C_{npp}(^4\text{He})$, obtained from GCF calculations using the SWV WS parameterization. The numerical results are presented by black squares. 
The blue and red symbols show the predictions of the scaling law, Eq. \eqref{eq:C3N_scaling}, with volume-plus-surface and volume-only contributions, respectively. The error bars are due to uncertainties in the fitted
parameters $L_V^{npp}, L_S^{npp}$, Table \ref{tab:npp_scaling}.
}
\label{fig:cppnSWV}
\end{figure}

\begin{figure}
\centering
\includegraphics[scale=0.55]{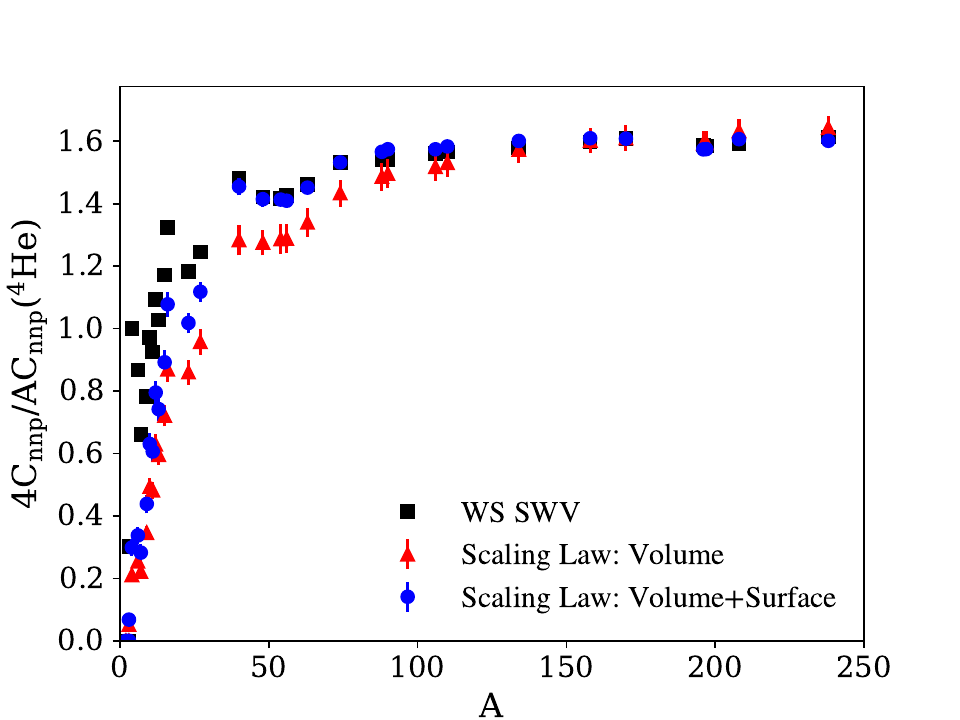}
\caption{
Mass-number dependence of the normalized $nnp$ three-body contact, 
$4 C_{nnp}(A)/A C_{nnp}(^4\text{He})$, obtained from GCF calculations using the SWV WS parameterization. The numerical results are presented by black squares. 
The blue and red symbols show the predictions of the scaling law, Eq. \eqref{eq:C3N_scaling}, with volume-plus-surface and volume-only contributions, respectively. The error bars are due to uncertainties in the fitted
parameters $L_V^{nnp}, L_S^{nnp}$, Table \ref{tab:nnp_scaling}.
}
\label{fig:cnnpSWV}
\end{figure}

\begin{figure}
\centering
\includegraphics[scale=0.55]{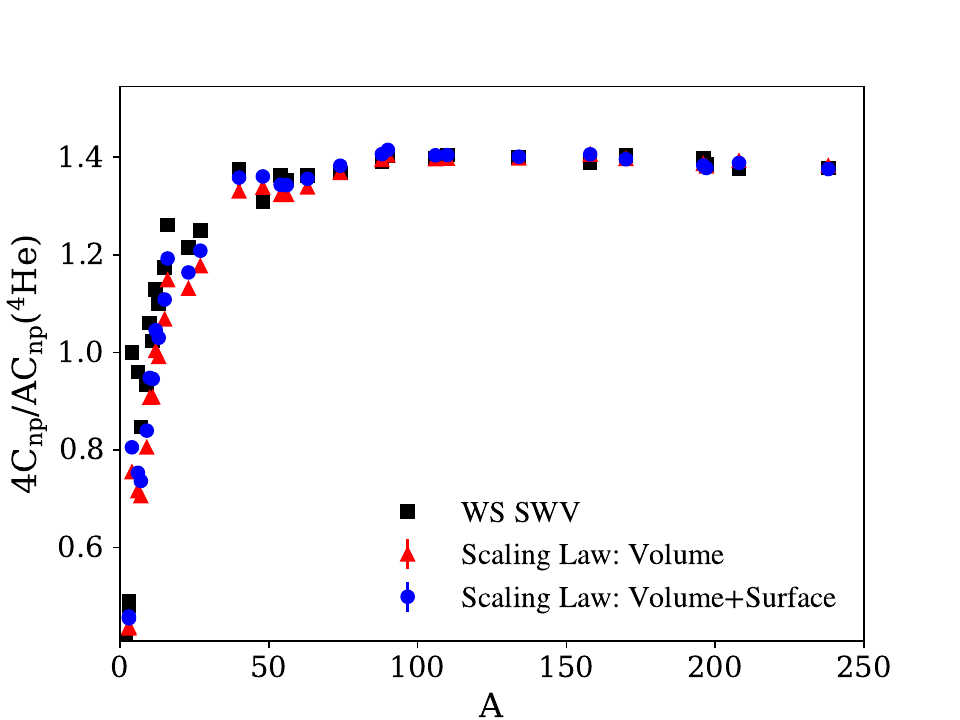}
\caption{
Mass-number dependence of the normalized $np$ spin-zero two-body contact, 
$4 C_{np}^{s=0}(A)/A C_{np}^{s=0}(^4\text{He})$, obtained from GCF calculations using the SWV WS parameterization. The numerical results are presented by black squares.  
The blue and red symbols show the predictions of the scaling law, Eq. \eqref{eq:C2N_scaling}, with volume-plus-surface and volume-only contributions, respectively. The error bars reflect the uncertainties in the fitted
parameters $L_V^{nnp}$, and $L_S^{nnp}$, Table \ref{tab:np_scaling}.
}
\label{fig:npSWV}
\end{figure}

Representative comparisons between the scaling relations and the
calculated contacts are shown in Figs.~\ref{fig:cppnSWV},
and \ref{fig:cnnpSWV} for the correlated triplets, and in Fig. \ref{fig:npSWV} for pairs. 
In all three figures the contacts are presented relative to the appropriate 
$^4$He contacts.
Inspecting these figures one can observe that for $A\geq 50$
the scaling law reproduces the
mean-field results remarkably well. The
agreement is particularly notable given that the model contains only a
volume and a surface contribution. As expected, the surface term is
most important for light nuclei and becomes progressively less relevant
with increasing mass number.

The fitted coefficients are listed in 
Tables~\ref{tab:npp_scaling}-\ref{tab:np_scaling}. Although the UR
and SWV parameterizations lead to different numerical values of the fit
parameters, both yield the same qualitative picture. The contacts are
dominated by the volume contribution, while the surface term acts as a
finite-size correction. The differences between the two
parameterizations are largely driven by the treatment of the lightest
nuclei, especially $^4$He, which serves as the normalization point for
the contacts.

A particularly interesting feature of the fitted coefficients is their
approximate isospin symmetry. Combining the UR and SWV results yields

\begin{align}
L_V^{npp} &= (1.9\pm0.3)\times10^{-3},\
L_V^{nnp} &= (2.1\pm0.1)\times10^{-3},
\end{align}

and

\begin{align}
L_S^{npp} &= (4.3\pm1)\times10^{-3},\
L_S^{nnp} &= (3.7\pm0.7)\times10^{-3}.
\end{align}

Within the quoted uncertainties, the volume and surface coefficients
for the $npp$ and $nnp$ channels are consistent with one another. This
agreement indicates that the underlying short-range dynamics remain
approximately isospin symmetric and that the observed differences
between the $npp$ and $nnp$ contacts originate primarily from the
different proton and neutron density distributions entering the scaling
relation rather than from the short-range coefficients themselves.

A similar pattern is observed for the spin-zero two-body contacts,
shown in Table~\ref{tab:scaling_params_2b}. 
We note that the extracted two-body coefficients exhibit 
smaller model dependence than the three-body, 
at the level of approximately $1\%$. The
agreement between the two- and three-body sectors suggests that the
same geometric mechanism governs the scaling of SRC pairs and
triplets.

\begin{table}[h]
\centering
\caption{Parameters for the scaling law fit of the three-body contact
$ C_{npp}$, Eq. \eqref{eq:C3N_scaling} normalized by $C_{npp}(^4\text{He})$.
The results are presented for two WS parameterizations - UR and SWV.
}
\label{tab:npp_scaling}
\begin{ruledtabular}
\begin{tabular}{lcc}
 & UR & SWV \\
\hline
$L_V^{npp}$ & $ 3.7\pm 0.2$ & $ 1.48 \pm 0.06 $ \\
$L_S^{npp}$ & $ 11 \pm 1$ & $ 2.7 \pm 0.3 $ \\
$\text{cov}(L_V^{npp},L_S^{npp})$  & $0.2$ & $-0.02$\\
$L_V^{abc}(L_S=0)$ & $5.63 \pm 0.05$ & $ 2.00 \pm 0.02 $
\end{tabular}
\end{ruledtabular}
\end{table}

\begin{table}[h]
\centering
\caption{Parameters for the scaling law fit of the three-body contact
$ C_{nnp}$, Eq. \eqref{eq:C3N_scaling} normalized by $C_{nnp}(^4\text{He})$.
The results are presented for two WS parameterizations - UR and SWV.
}
\label{tab:nnp_scaling}
\begin{ruledtabular}
\begin{tabular}{lcc}
 & UR & SWV \\
\hline
$L_V^{nnp}$ & $ 4.3\pm 0.1$ & $ 1.59 \pm 0.05 $ \\
$L_S^{nnp}$ & $ 8.7 \pm 0.7$ & $ 2.5 \pm 0.2 $ \\
$\text{cov}(L_V^{nnp},L_S^{nnp})$  & $-0.09$ & $-0.01$\\
$L_V^{abc}(L_S=0)$ & $5.94 \pm 0.04$ & $ 2.06 \pm 0.01 $
\end{tabular}
\end{ruledtabular}
\end{table}

\begin{table}[h]
\centering
\caption{Parameters for the scaling law fit of the spin zero two-body contacts
$ C_{np}^{s=0}$, Eq. \eqref{eq:C2N_scaling} normalized by $C_{np}^{s=0}(^4\text{He})$.
The results are presented for two WS parameterizations - UR and SWV.
}
\label{tab:np_scaling}
\begin{ruledtabular}
\begin{tabular}{lcc}
 & UR & SWV \\
\hline
$L_V^{np}$ & $ 1.78\pm 0.03$ & $ 1.14 \pm 0.01 $ \\
$L_S^{np}$ & $ 1.1 \pm 0.2$ & $ 0.23 \pm 0.07 $ \\
$\text{cov}(L_V^{np},L_S^{np})$  & $-0.2$ & $-1\times 10^{-3}$\\
$L_V^{abc}(L_S=0)$ & $1.99 \pm 0.01$ & $ 1.184 \pm 0.002 $
\end{tabular}
\end{ruledtabular}
\end{table}

\par

\begin{table}[h]
\centering
\caption{
Parameters for the scaling law fit of the spin zero two-body
$nn$, $np$, and $pp$ contacts, Eq. \eqref{eq:C2N_scaling}.
The results are presented for an average of the two WS parameterizations - UR and SWV.
}
\label{tab:scaling_params_2b}
\begin{ruledtabular}
\begin{tabular}{lccc}
 & $np$ & $nn$ & $pp$ \\
\hline
$L_V \times 10^{2}$ & $2.2\pm 0.2$ & $2.3 \pm 0.01$ & $2.4\pm 0.01$ \\
$L_S \times 10^{2}$ & $0.9 \pm 0.8$ & $1.0 \pm 0.03$ & $0.79\pm 0.06$ \\
\end{tabular}
\end{ruledtabular}
\end{table}
\section{Discussions and Conclusions} 
We have extended the generalized contact formalism to three-nucleon short-range correlations and developed a practical framework for calculating three-body nuclear contacts using mean-field wave functions. The resulting contacts exhibit a simple and systematic dependence on nuclear mass and composition across the nuclear chart.

Motivated by the approximately linear growth of the calculated contacts with mass number, we derived a phenomenological scaling relation consisting of a volume term and a finite-size surface correction. Despite its simplicity, the scaling relation reproduces the calculated three-body contacts with good accuracy for both Woods--Saxon parameterizations considered in this work. The surface contribution is most important for light nuclei, while the volume term dominates for heavier systems.

We also find that the $npp$ and $nnp$ contacts exhibit a systematic splitting in neutron-rich nuclei. This behavior closely resembles that previously observed for spin-zero two-body contacts and can be understood as a consequence of the different neutron and proton density distributions. The corresponding contact ratio follows a nearly linear dependence on neutron excess and is quantitatively similar to the two-body case.

Our results suggest that three-body contacts obey universal scaling patterns analogous to those governing short-range correlated pairs. Future work should test these predictions using \emph{ab initio} wave functions and compare them with forthcoming experimental studies of 3N-SRCs.

\section{Acknowledgement} 

This work was supported by the Pazy Foundation, grant number 443/2021. 


\end{document}